\documentclass{cernrep} 
\usepackage{texnames}
\usepackage[T1]{fontenc}
\begin{document}
\title{One-nucleon transfer reactions and the optical potential}
 
\author{F.M. Nunes$^{1,2}$, A. Lovell$^{1,2}$, A. Ross$^{1,2}$,  L.J. Titus$^{1,2}$, R.J. Charity$^{3}$, W.H. Dickhoff$^{4}$, M.H. Mahzoon$^{4}$, J. Sarich$^5$, S. M. Wild$^5$ }

\institute{$^{1}$National Superconducting Cyclotron Laboratory, Michigan State University, USA\\
$^{2}$Department of Physics and Astronomy, Michigan State University, USA\\
$^{3}$Department of Chemistry, Washington University, Saint Louis, USA\\
$^{4}$Department of Physics, Washington University, Saint Louis, USA\\
$^{5}$Mathematics and Computer Science Division, Argonne National Laboratory,  Illinois, USA
}

\maketitle 

\begin{abstract}
We provide a summary of  new developments in the area of direct reaction theory with a particular focus on one-nucleon transfer reactions. We provide a status of the methods available for describing (d,p) reactions. We  discuss the effects of nonlocality in the optical potential in transfer reactions. The results of a purely phenomenological potential and the optical potential obtained from the dispersive optical model are compared; both point toward the importance of including nonlocality in transfer reactions explicitly. 
Given the large ambiguities associated with optical potentials, we discuss some new developments toward the quantification of this uncertainty. We conclude with some general comments and a brief account of new advances that are in the pipeline.     
\end{abstract}

\section{Introduction}
Nuclear reactions are an important and versatile tool to study nuclei, particularly those nuclei away from stability. One-nucleon transfer reactions can
provide information about the single-particle structure of the nucleus of
interest (see e.g., \cite{exp1,exp2}).  Deuterons are often used as a probe. However, due to its loosely bound
nature, it is understood that deuteron breakup effects need to be considered carefully.

A number of experimental programs in rare-isotope facilities worldwide are also using (d,p) and (d,n) reactions as a tool to extract  capture rates of astrophysical relevance (e.g., \cite{exp3,exp4}). 
Given the many ongoing and planned experimental efforts in this direction, it is critical that the reaction theory used to interpret those results be reliable and that the uncertainties in the reaction theory be well understood.

\section{Status of the treatment of  reaction dynamics}

The treatment of deuteron induced reactions involving intermediate and heavy nuclei pose severe challenges to microscopic theory.
For most reactions involving these systems, the problem is cast as a
three-body scattering problem with effective nucleon-nucleus interactions. We  first briefly review the current status of reaction theories for describing
deuteron-induced one-nucleon transfer reactions.

The continuum discretized coupled channel method (CDCC) \cite{cdcc} developed in the eighties has now been benchmarked against the exact Faddeev method \cite{fadd}. The results of this benchmark \cite{upadhyay2012} demonstrate that while the reduction to one Jacobi component is adequate for transfer reaction at low energies, the convergence rate for the low-energy breakup distributions is extremely slow and does not allow for reliable extrapolations. Reaction calculations at higher energies introduce the issue of the energy dependence in the optical potentials. While in coordinate-based methods there is typically a fixed choice of the energy at which the optical potentials are evaluated, in the current implementation of the Faddeev method \cite{fadd}, calculations are performed in momentum space and can take into account the explicit energy dependence of the interaction in the reaction dynamics. This poses an ambiguity in the comparison that is discussed in detail in \cite{upadhyay2012}.

 While both the CDCC and Faddeev methods are computationally intensive, the adiabatic distorted wave approximation (ADWA) \cite{adwa}, which is essentially a simplification of CDCC to make all the channels in the continuum degenerate with the ground state (adiabatic approximation), provides a very efficient tool to analyze transfer reactions and explicitly includes deuteron breakup to all orders. A benchmark of this approach with the exact Faddeev has also been performed \cite{nunes2011} and the results show that the adiabatic method, while only valid for deuteron-induced transfer reactions, provides as good an agreement as the CDCC method. 
 
Understanding the regions of validity of these reaction theories is very important. However, the current implementation of the Faddeev method  \cite{fadd} also has its limitations, a fact that became clear during the benchmarking process \cite{nunes2012}. As the effects of the Coulomb force  increase, namely as the energy decreases and/or the charge increases, the method of Coulomb screening used by \cite{fadd} begins to fail. For example, currently we are not able to produce exact cross sections for $^{10}$Be(d,p) below $\approx 5$ MeV, $^{48}$Ca(d,p) below $\approx 15$ MeV,
 or any reaction cross section of interest for isotopes heavier than Ni. One way to overcome this problem is to cast the Faddeev equations in the Coulomb basis instead of the traditional plane wave basis \cite{elster}. By doing this, one avoids Coulomb screening completely.
The TORUS collaboration \cite{torus}, a topical collaboration in nuclear theory with the goal of advancing methods of (d,p) reactions,
has put in place many pieces of the puzzle necessary to solve the Faddeev equations in the Coulomb momentum space basis \cite{upadhyay2014,separable,vasily}.
We expect in the next couple of years to have a completely new and fully developed code that can address the three-body problem of A(d,p)B for nuclei with large $Z$.

\section{Nonlocality in optical potentials and transfer reactions} 

One of the most important ingredients in predictions of reaction cross sections are the optical potentials.
From a microscopic point of view, it is clear that the nucleon-nucleus optical potential should be non local due to many-body effects, including antisymmetrization and excitations. However, phenomenological potentials have traditionally been assumed local, for convenience. The consequence of this simplification is that it becomes strongly energy dependent. As many-body methods improve, including more and more correlations and a careful treatment of the continuum, one can expect the optical potential to be fully extracted from these many-body theories in the near future.  It is therefore timely for reaction theory to be prepared to handle nonlocality in the effective interactions and to understand the magnitude of the effects.

\begin{figure}
\centering\includegraphics[width=.8\linewidth]{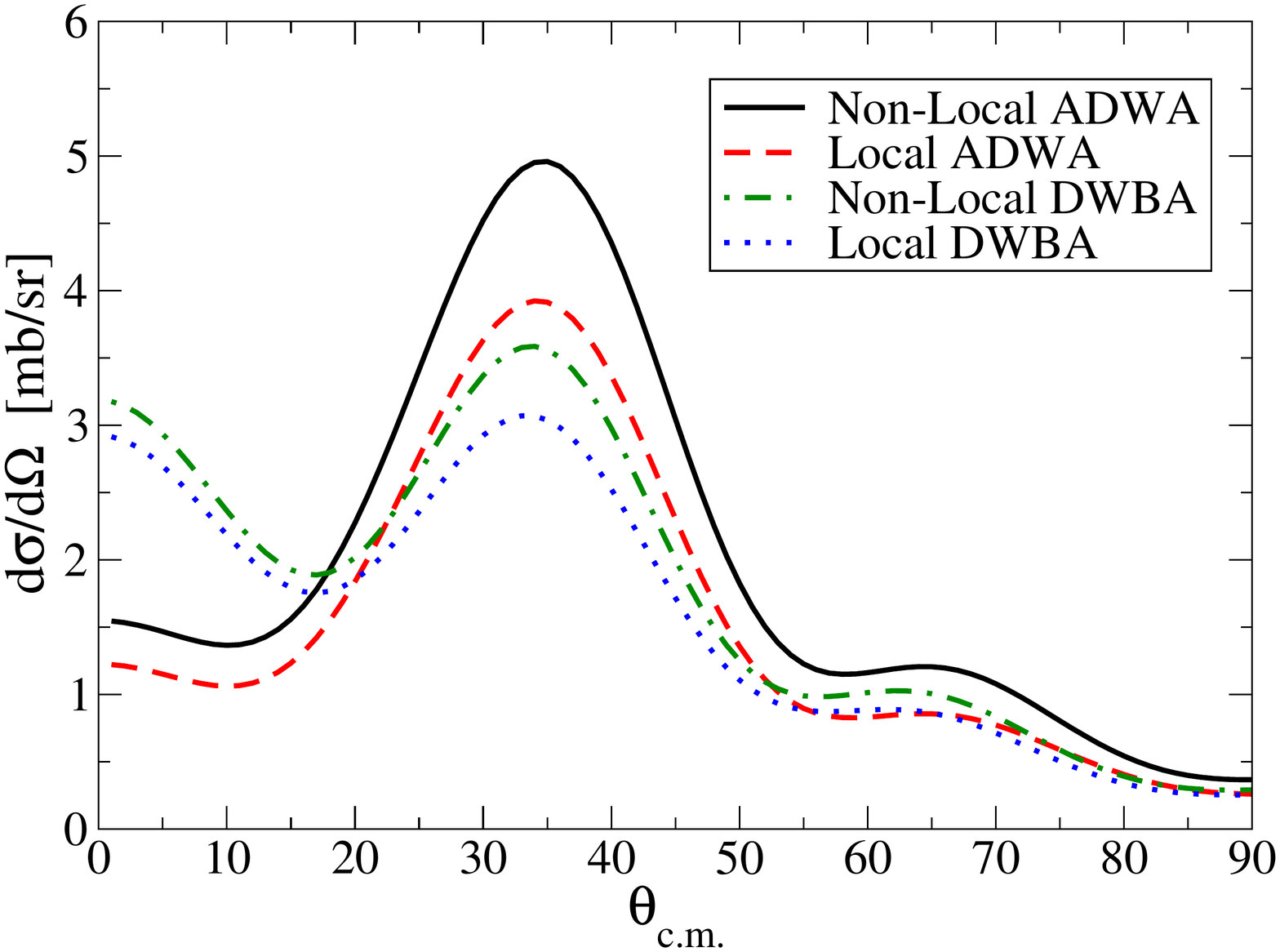}
\caption{Angular distributions for $^{126}$Sn(d,p)$^{127}$Sn at $E_d=20$ MeV: comparing the inclusions of nonlocality in the exit channel for ADWA and for DWBA. }
\label{fig:1}
\end{figure}

\begin{figure}
\centering\includegraphics[width=.8\linewidth]{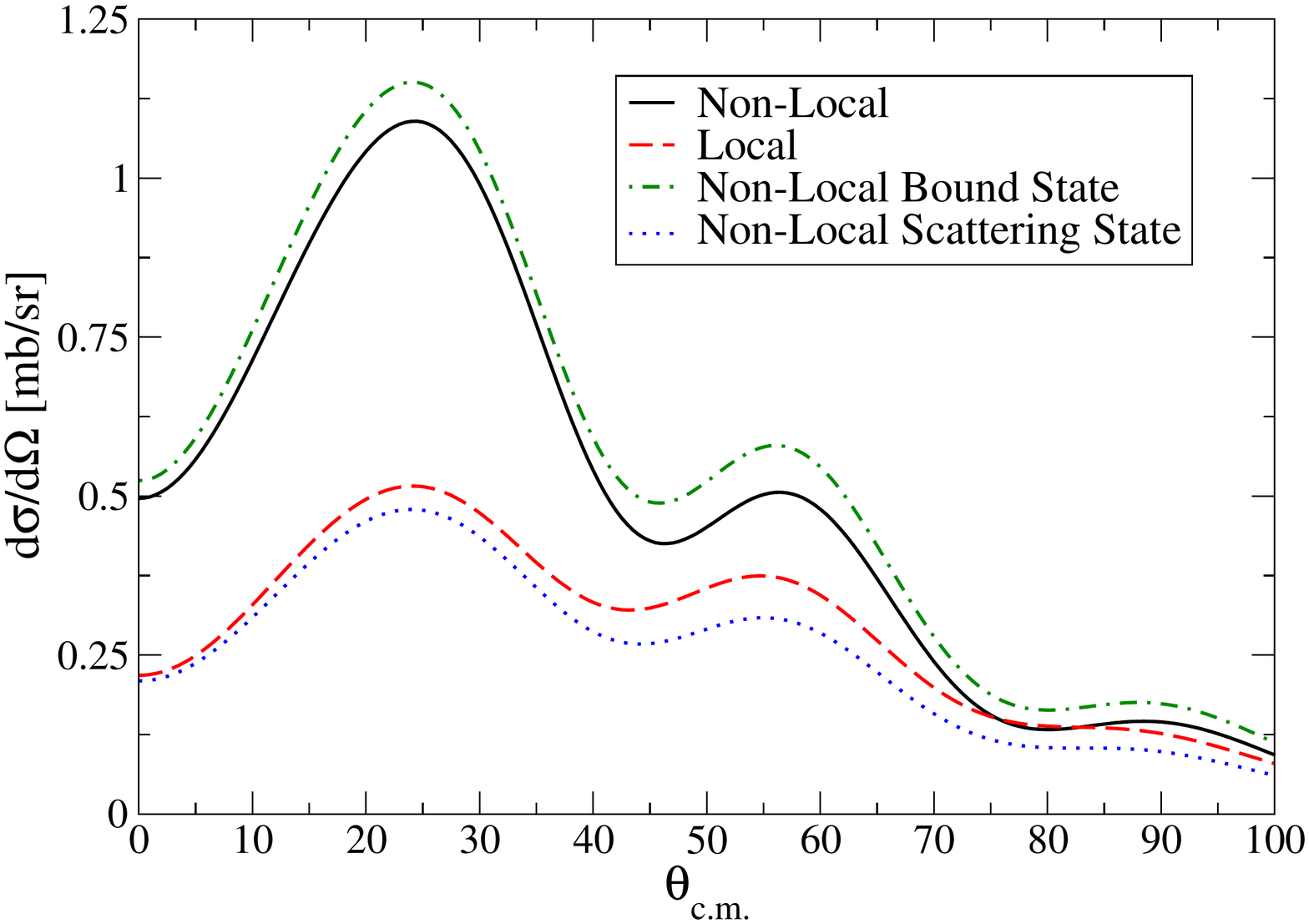}
\caption{Angular distributions for $^{126}$Sn(d,n)$^{127}$Sb at $E_d=20$ MeV: comparing the inclusions of nonlocality in the exit channel for ADWA and for DWBA. }
\label{fig:2}
\end{figure}

It is clear that for any non-local potential, one can construct local phase-equivalent potentials, i.e., reproducing the exact same elastic scattering. Even when the elastic scattering is exactly reproduced, nonlocality can imprint itself in other channels. In \cite{titus2014,ross2015} the problem is addressed for one-nucleon transfer reactions. In the first of these works \cite{titus2014}, a large number of reactions at different beam energies and for different targets are considered. The Perey and Buck non-local optical potential was used \cite{pb} and local phase-equivalent potentials were constructed. Finally, the inclusion of nonlocality in one-nucleon transfer reactions populating single-particle states was studied. The study was performed in DWBA and nonlocality was included only in the exit channel, namely in the neutron-bound state and the proton-distorted wave. A similar study was performed using the dispersive optical model \cite{dom}. Generally, the nonlocality in the scattering state reduced the cross section, while the nonlocality in the bound state increased it, with the net effect being an increase as large as $30$\%. In \cite{ross2015} we consider hole states in $^{40}$Ca, again using DWBA and nonlocality only in the exit channel. The dispersive optical model and the Perey and Buck potentials provide similar results. The effect of nonlocality on transfer is very large for these hole states (up to $50$\%).

In Fig. 1  we show the predicted transfer cross sections for $^{126}$Sn(d,p)$^{127}$Sn at $E_d=20$ MeV. Comparing the local DWBA predictions (blue dotted line) with the corresponding non local (green dot-dashed line), we see an effect of up to $20$\% in magnitude. Also shown, for completeness, are the same results within ADWA (red dashed for the local and black solid for the non local). While ADWA and DWBA show noticeably different angular distributions (a difference coming from the different treatment of deuteron breakup), the relative effect of nonlocality  is magnified in ADWA. Note  that in these ADWA calculations, the nonlocality in the deuteron channel is not yet included. Nonlocality in the deuteron channel will be discussed elsewhere \cite{titus2015}.  

Because nonlocality is introduced in the nuclear interaction only, one would expect to find similar effects in the study of (d,n) reactions.
In Fig. 2 we show the angular distributions for $^{126}$Sn(d,n)$^{127}$Sb at $E_d=20$ MeV. All lines in Fig. 2 correspond to DWBA calculations. The result when using the local interactions is shown by the red dashed line and that including the non-local interactions is the black solid line. Note that the ground state of $^{127}$Sb is different than that of $^{127}$Sn, given the different proton and neutron numbers of the target. While the ground state of $^{127}$Sn is a $1h_{11/2}$ state,  in $^{127}$Sb the ground state is $1g_{7/2}$ with a significant difference in the binding energy. In addition, the proton state is subject to Coulomb repulsion. All these differences compound to produce a very different effect of nonlocality for the (d,p) and (d,n) reactions on heavy nuclei. For $^{126}$Sn(d,n)$^{127}$Sb, the effect of nonlocality is very large, doubling the cross section at the peak. We also show in Fig. 2 the separate effects of introducing nonlocality in the neutron bound state (green dot-dashed line) and the proton scattering state (blue dotted line). Nonlocality in the  proton scattering state reduces the cross section but only by a little. The larger effect is in the nonlocality in the bound state.

These studies all call for the need to include nonlocality explicitly in reactions for a reliable description of the process.

\section{Optical potential uncertainties}

\begin{figure}
\centering\includegraphics[width=.6\linewidth]{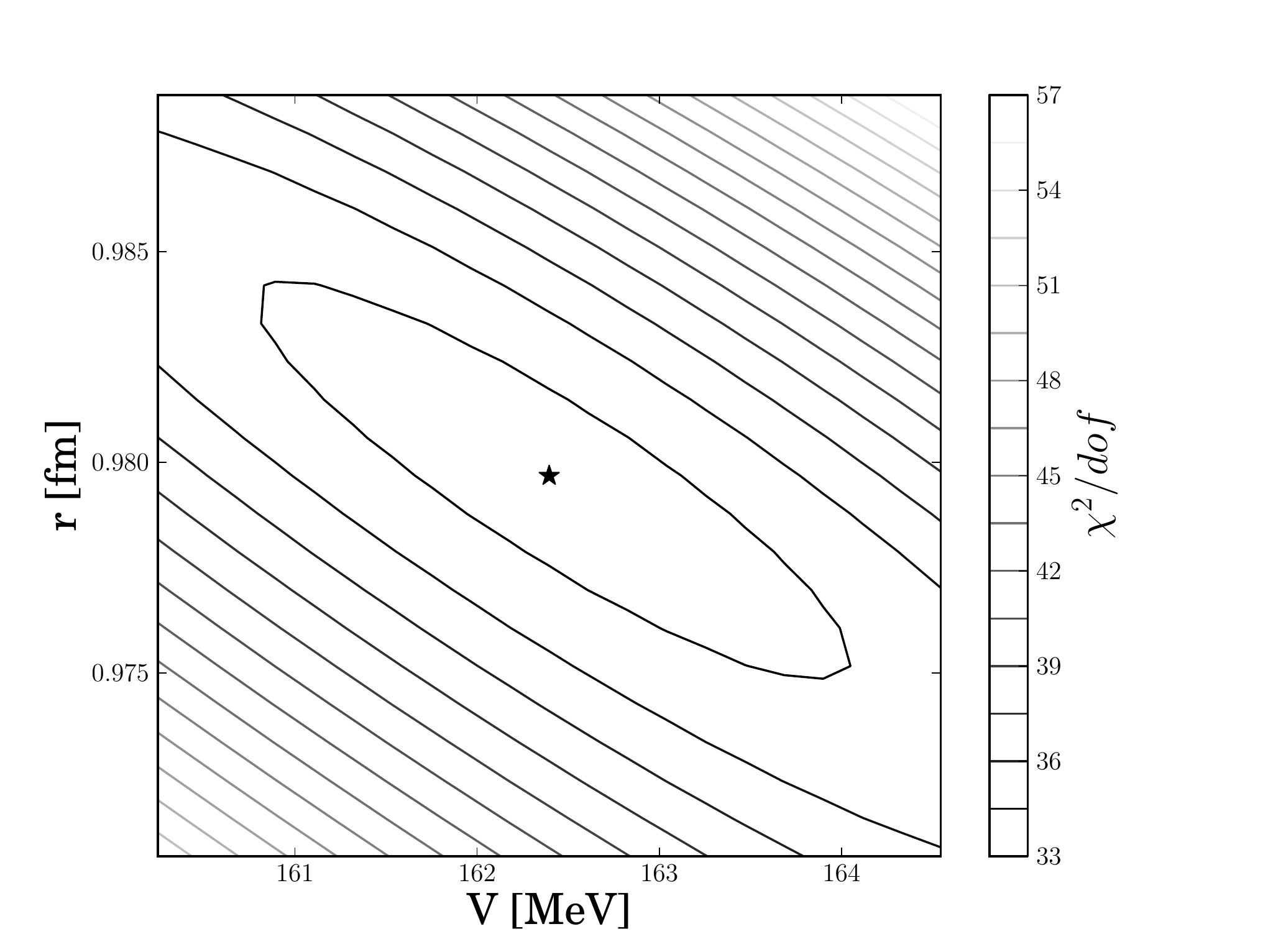}
\caption{Contour plot for the residuals resulting for the fit of neutron elastic scattering on $^{48}$Ca at 23 MeV, when varying the radius parameter $r$ of the real part of the optical potential and the corresponding depth $V$.}
\label{fig:3}
\end{figure}

\begin{table}[b!]
\begin{center}
\begin{tabular}{| c | c | c | c |}
\hline & Depth [MeV] & Radius [fm] & Diffuseness [fm] \\ \hline
Volume Term (real) & 162.4 (103.2) & 0.98 (1.05) & 0.69 (0.86) \\ \hline
Surface Term (imaginary) & 48.7 (16.5) & 1.1 (1.43) & 0.30 (0.67) \\ \hline
Spin-Orbit Term & 3.33 & 1.07 & 0.66 \\ \hline
\end{tabular}
\label{bestfit}
\caption{Best fit parameters for the elastic scattering of $^{48}$Ca(d,d)$^{48}$Ca. The spin-orbit term was not fit. All optical model potential parameters not listed here were zero. The optimization was initialized at the value in parentheses.}
\end{center}
\end{table}

The study in \cite{exp1} demonstrates that the optical potential uncertainties are reduced in ADWA when compared to DWBA, the main reason being that ADWA relies only on nucleon optical potentials, whereas DWBA requires a deuteron optical potential that is less well constrained. Knowing that these effective interactions carry ambiguities, it is  important to quantify the uncertainties in the predicted reaction observables. So far estimates of these uncertainties have been performed by comparing the results when two arbitrary optical potentials are chosen (e.g., \cite{exp2}). Then the results of the estimated error bars depend on the choice of the two representative potentials. Alternatively, one can make use of modern techniques in the field of uncertainty quantification and explore these tools in the domain of reaction theory.

 Our plan is to use modern statistical tools to quantify uncertainties in nuclear reactions (see e.g., \cite{wild2015}). This is largely an untapped field and therefore we start with a simple case.  
We look at the case of $^{48}$Ca(d,p)$^{49}$Ca in DWBA  and examine the effect of the uncertainty from constraining the deuteron optical potential on the predicted transfer cross section. We use elastic scattering data at 23.2 MeV from \cite{ca48-data} and perform a $\chi^2$ minimization to find the best fit. We take a typical Woods-Saxon shape and the parameterization of \cite{ca48-start} as a starting point for the minimization routine. We allow all three parameters in the central volume real and the central surface imaginary terms (depth, radius, and diffuseness)  to vary simultaneously (the spin-orbit interaction is kept fixed).  Parameters are only allowed to vary within physical limits. The minimum found is summarized in Table I, with initial parameters shown in parentheses. We then verify that the distribution of the total $\chi^2$ in the elastic channel, around the minimum, is Gaussian. This can be seen by the elliptical behavior of the 
$||\chi^2||$ contour plots. In Fig. \ref{fig:3} we show an example of such a plot, where the vertical axis is the potential depth and the horizontal axis is the radius of the volume term.

\begin{figure}
\centering\includegraphics[width=.7\linewidth]{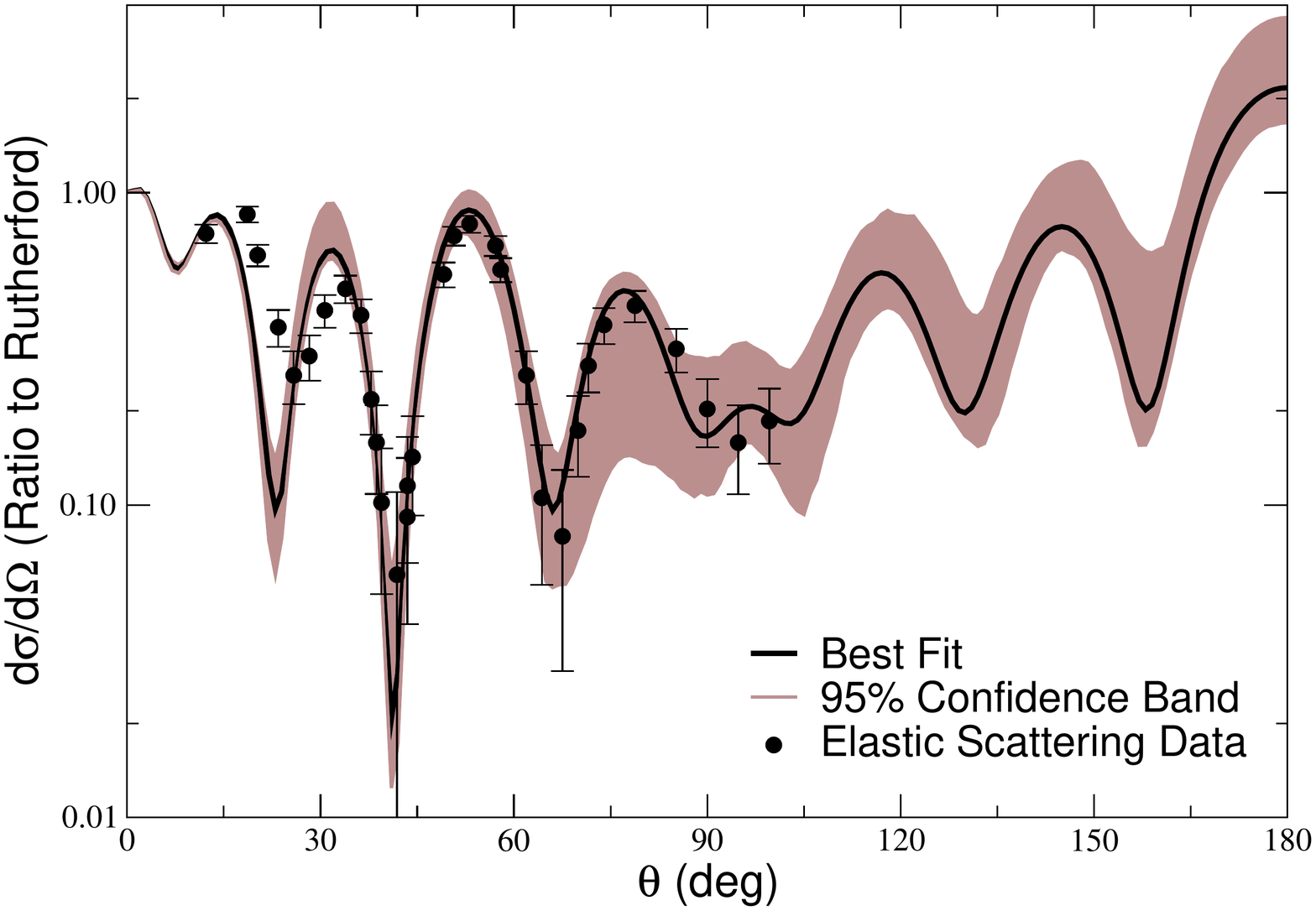}
\caption{Elastic scattering angular distribution for neutrons on $^{48}$Ca at 23.2 MeV: solid black line is the best fit, the brown corresponds to the 95\% confidence band and in black are the data points \cite{ca48-data}.}
\label{fig:4}
\end{figure}
\begin{figure}
\centering\includegraphics[width=.7\linewidth]{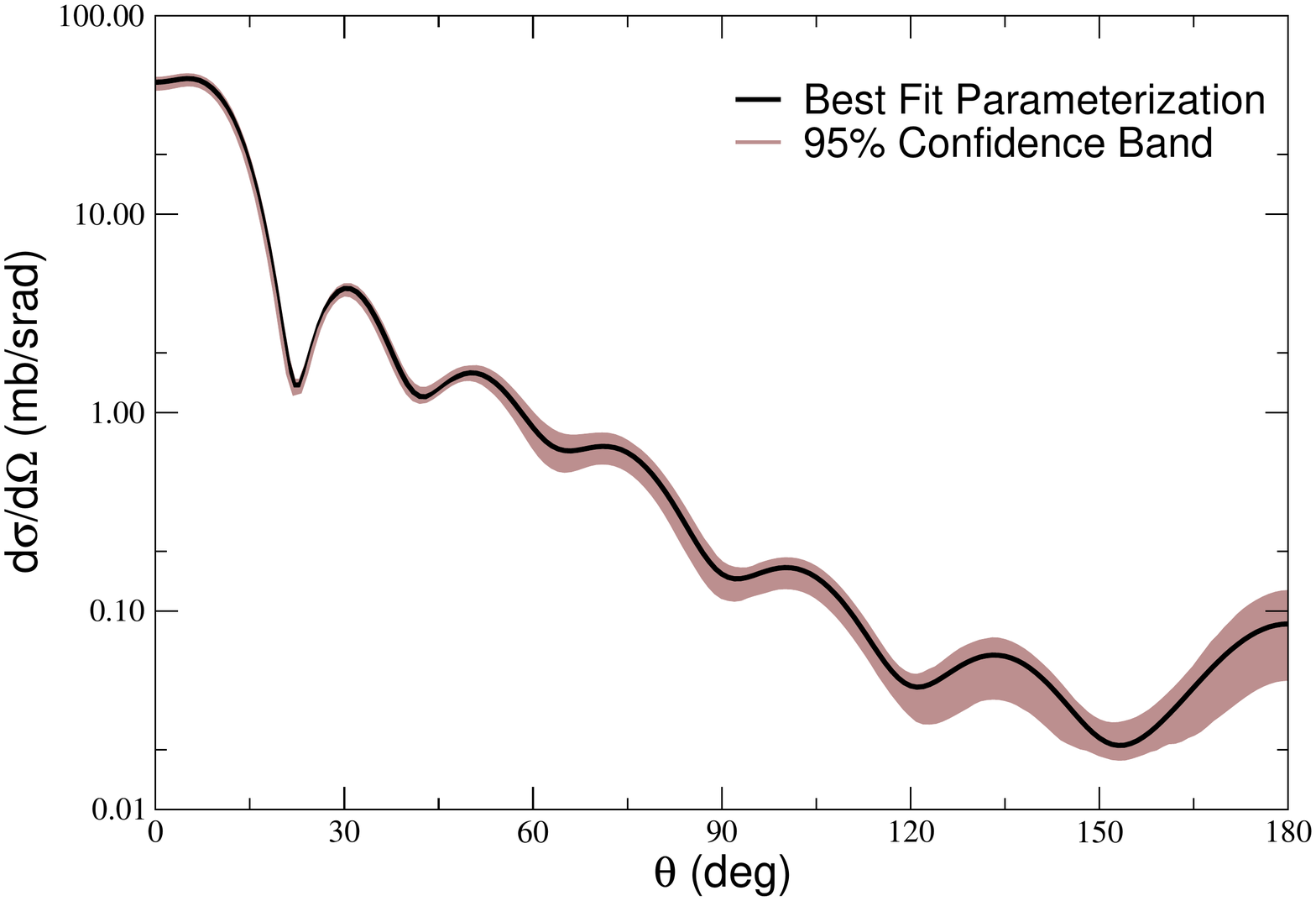}
\caption{Angular distribution for  $^{48}$Ca(d,p)$^{49}$Ca at 23.2 MeV: solid black line is the best fit, the brown corresponds to the 95\% confidence band.}
\label{fig:5}
\end{figure}
We then pull 200 sets randomly from the Gaussian distribution based on the elliptical behaviour surrounding the minimum, throw out the largest 5 and lowest 5 predicted cross sections per angle, to obtain  the $95$\% confidence band. Using the same 200 sets, we define the $95$\% confidence bands independently for elastic and transfer. The results are depicted in Figs. \ref{fig:4} and \ref{fig:5}.

The error bars considered in quantifying the $\chi^2$ for these calculations were the experimental error bars. However, often the difference between data and theory is larger than the experimental error bars. One can then repeat the procedure including the theoretical error.
Also noteworthy, there are cases in which the distribution around the minimum is not Gaussian. Then one may need to pull the random sets from the regions of constant $\chi^2$  directly and/or appeal to Bayesian techniques \cite{higdon,mcdonnell}.

A refinement of the fitting can be done by including a larger set of data, with perhaps a variety of observables. As an example, we have started to look at simultaneously fitting elastic and inelastic cross sections, including target excitation in the reaction model. While a larger number of parameters are available for fitting, the actual procedure for extracting confidence bands is the same. A more challenging problem is that of the uncertainty in the theoretical model itself. More work will be needed to determine the best approach.

\section{Outlook}

As described here, there are a number of exciting developments in reaction theory. The implementation of the Faddeev equations in the Coulomb basis is progressing and will provide reliable predictions for (d,p) reactions for a wide range of nuclei and energies.
The studies performed on non-local interactions demonstrate the need to include these interactions explicitly in describing transfer processes.
While these are critical studies for the future of the field, they still rely on the knowledge of the effective interactions between nucleons and the target.

Although traditionally these effective interactions have been extracted phenomenologically, we are currently beginning an era where the level of nuclear many-body methods may enable the extraction of these potentials directly from the NN interaction.  A new collaboration between our group and the ORNL theory group aims at extracting these potentials from coupled-cluster theory. While one might still expect a deficiency in the absorptive component, the inclusion of a variety of correlations and the continuum basis hold promise for using this method versus other many-body methods.

Last but not  least, we are starting to develop techniques to quantify uncertainties in reaction theory. This work is still in its infancy. As described in this proceedings, we are still exploring the best way to characterize the uncertainties for the limited problem of the effect of the errors in a given data set in constraining the interaction and the propagation to reaction observables. There are many other sources of uncertainties in reaction theory, which may be equally or even more important. These need to be considered case by case and then accounted for jointly so that we can provide a reliable and useful procedure for  uncertainty quantification in this field.

\vspace{0.5cm}
This work was supported by the National Science
Foundation under Grant No. PHY-1403906  and PHY-1304242 and the Department
of Energy, Office of Science, Office of Nuclear Physics under award No. DE-FG52-08NA28552, DE-SC0004087, DE-87ER-40316.
This work was also supported by the U.S. Department of Energy, Office of Science,
Advanced Scientific Computing Research program under contract number DE-AC02-06CH11357.

\end{document}